\documentstyle[11pt,twoside,sympo2011,epsfig]{article}
\begin{document}
\title{HD 49933: A laboratory for magnetic activity cycles}

\author{T. Ceillier$^{*,1}$, J. Ballot$^{2,3}$,  R.A. Garc\'\i a$^{1}$, G.R. Davies$^1$, S. Mathur$^4$, T.S. Metcalfe$^4$, D. Salabert$^5$
}
\affil{$^*$ Affiliations are listed at the end of the paper} 
%
\begin{abstract}
Seismic analyses of the CoRoT target HD~49933 have revealed a magnetic cycle. Further insight reveals that frequency shifts of oscillation modes vary as a function of frequency, following a similar pattern to that found in the Sun. In this preliminary work, we use seismic constraint to compute structure models of HD~49933 with the Asteroseismic Modeling Portal (AMP) and the CESAM code. We use these models to study the effects of sound-speed perturbations in near surface layers on p-mode frequencies.
\end{abstract}
\section{Introduction and observations}
Dynamo processes in the outer convective envelope of solar-like stars generate magnetic fields, which create active regions and spots at the stellar surface (e.g. Lanza 2010). In the Sun, such a dynamo leads to a well known almost regular 22-year magnetic cycle. 
This dynamo is far from being completely understood, and is difficult to predict as suggested by the recent unusually long solar minimum of cycle 23 (e.g. Salabert et al. 2009). 
Study of magnetic cycles in other stars should allow us to better understand the physical mechanisms by studying many stars in different evolutionary stages and conditions (e.g. Chaplin et al. 2007; Metcalfe et al. 2007).  
Variations of magnetic fields modify the stellar structure and therefore frequencies and amplitudes of the acoustic modes (Jim\'enez-Reyes et al. 2003). Such variations are detectable with current asteroseismological facilities.

HD 49933 is a F5V star observed by CoRoT for 60 and 137 days in 2007 and 2008 (Appourchaux et al. 2008; Benomar et al. 2009). The light curve presents clear signatures of active regions crossing the visible stellar disk that reveal a rotation period of 3.5 days with differential rotation. The analysis of short time series reveals frequency shifts and amplitude variations in acoustic modes unveiling short-term variations of activity, compatible with a short stellar cycle of around 150 days (Garc\'\i a et al. 2010). Moreover, the frequency shifts measured in HD~49933 present a clear dependence with increasing frequency, reaching a maximum shift of about 2 $\mu$Hz around 2100 $\mu$Hz (Salabert et al. 2011). Such a dependence is comparable to the one observed in the Sun, which is understood to arise from changes in the outer layers due to its magnetic activity.

\section{Modelling and Discussion}
In this work we study the internal structure of HD~49933 paying special attention to properties of the external convection zone. We use the seismic and spectroscopic observations to find a best model with AMP (Metcalfe et al. 2009). We use this model as a starting point to compute two structure models with the CESAM stellar evolution code (Morel 1997). The first model uses the mixing-length theory (MLT, B\"ohm-Vitense 1958) for treating convection whereas the second use the CGM prescription (Canuto et al 1996). To quantify the sensitivity of modes to the superficial structure, we compute upper turning points of the p modes with frequencies $\nu \in$ [1400--2600]~$\mu$Hz. Turning points $r_o$ are computed as $2\pi\nu=\omega_c(r_o)$, where $\omega_c=c_s/(2H)$ is the isothermal cut-off frequency ($c_s$ is the sound speed and $H$ the pressure scale height). In Fig.~1a, we see similar results for both models. Nevertheless, in the CGM model, $r_o$ depends less on $\nu$ in the range 1400--2600~$\mu$Hz, because the temperature profile obtained with this prescription is sharper than the one obtained with MLT.

We crudely estimate the impact on frequencies of a perturbation in pressure due to a change in magnetic field.
We assume that $c_s$ is perturbed by a change $\delta p$ in pressure, all other quantities staying unchanged.
Thus, the travel time of waves is modified by $\delta\tau =\int^{r_o} \delta p / (p c_s)\,\mathrm{d}r$. We then consider the effect on frequencies is $\delta\nu \approx \nu^2\delta\tau$.  We consider two profiles for $\delta p$: (A) $\delta p$ is constant in the surface layers; (B) $\delta p$ linearly grows during the first 10 Mm beneath the surface. Results are plotted in Fig. 1b. We use $\delta p = 400\:\mathrm{dyn\,cm^{-2}}$ at the photosphere to recover reasonable values of $\delta\nu$.

\begin{figure}[!ht]
\begin{center}
\epsfig{width=7.5cm,file=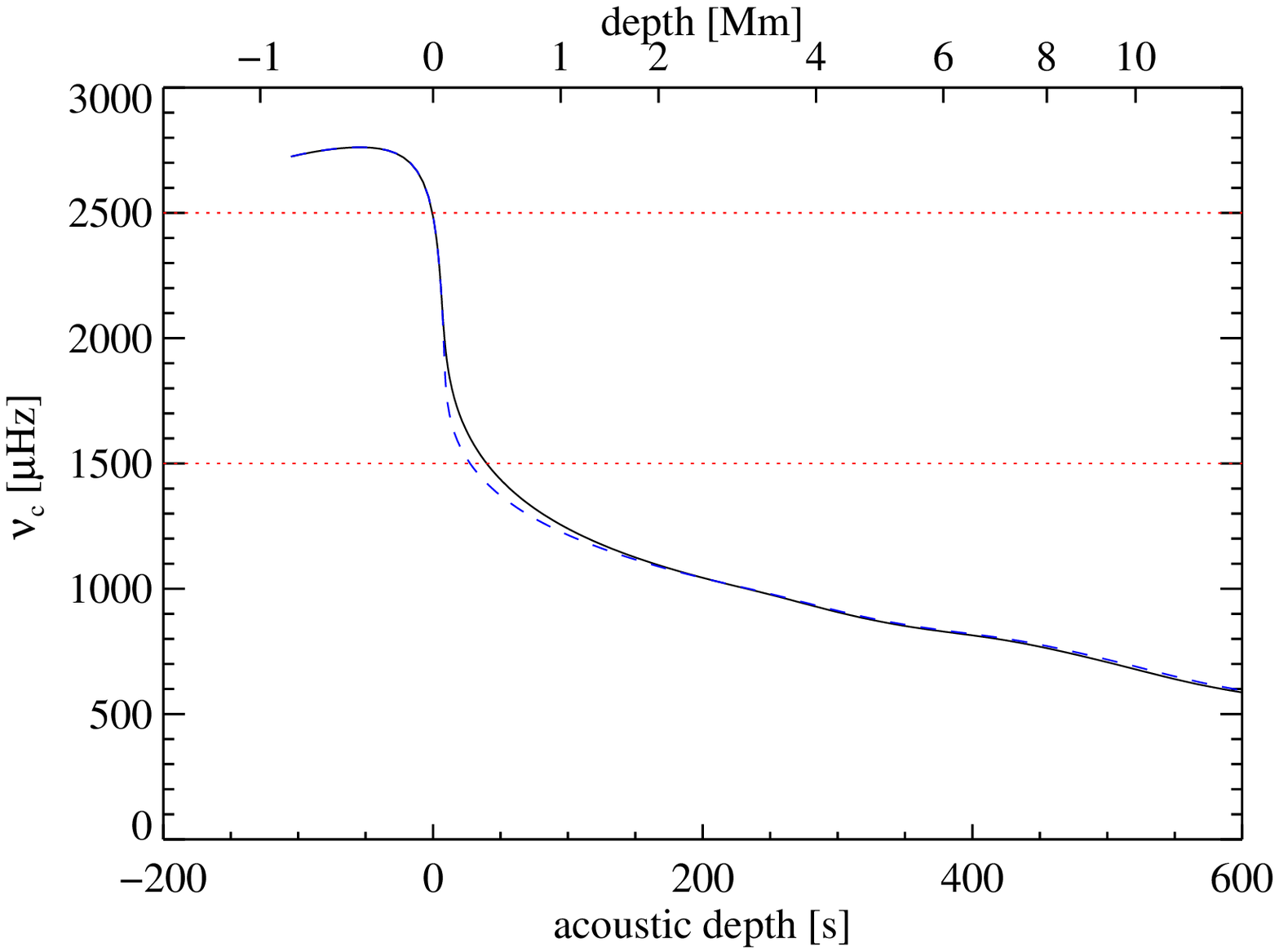}\epsfig{width=7.5cm,file=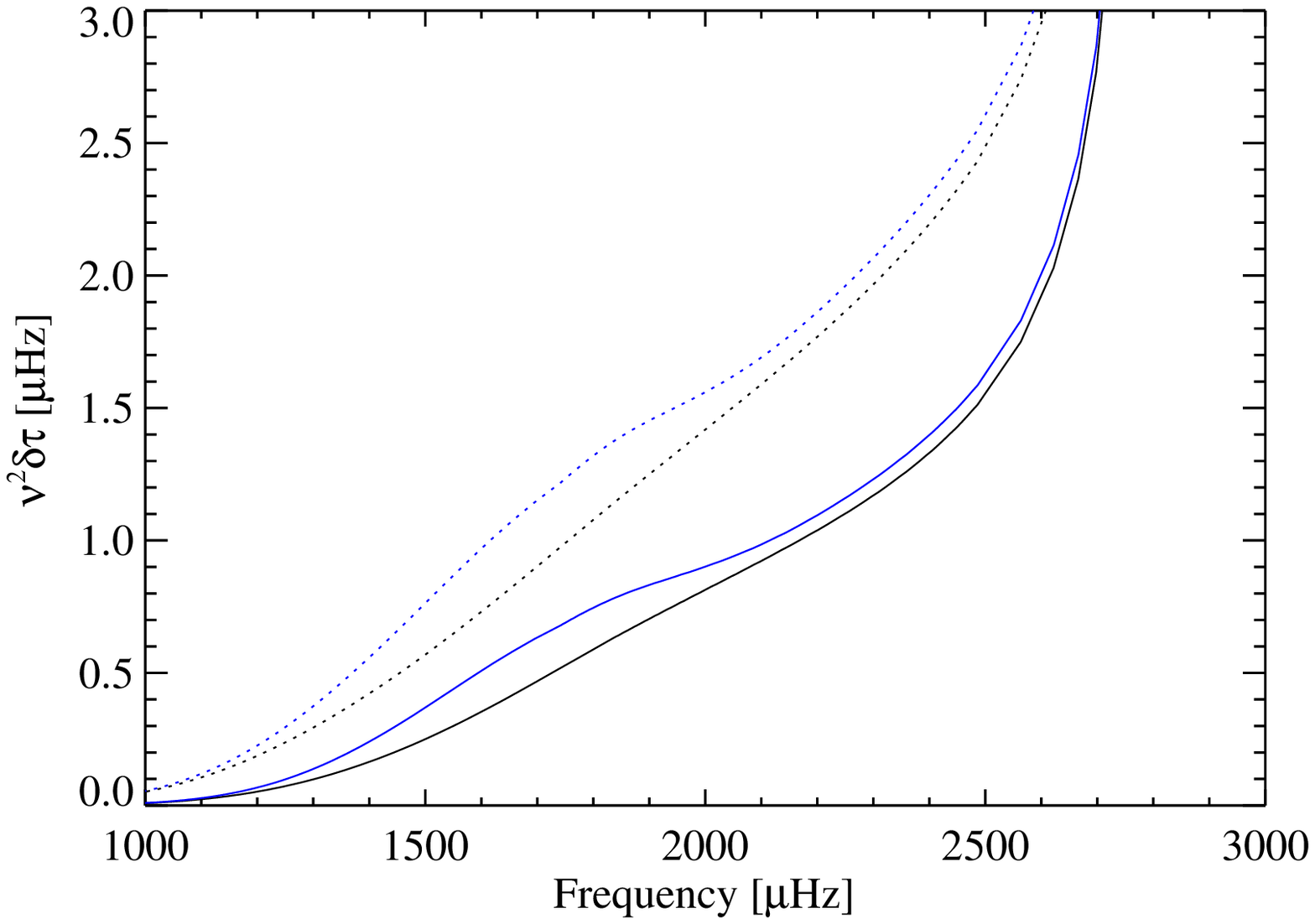}
\caption{\emph{a)} Acoustic cut-off frequency as a function of depth: MLT (solid black line) and CGM (blue dashed line) models of -- \emph{b)} Effects on frequencies of a pressure change following the profile A (solid lines) or B (dotted lines). Same color code as before.}
\end{center}
\end{figure}

\acknowledgments{The CoRoT space mission has been developed and is operated by CNES, with contributions from Austria, Belgium, Brazil, ESA (RSSD and Science Program), Germany and Spain. JB, RAG, and DS acknowledge the support given by the French PNPS program and CNES. NCAR is supported by the National Science Foundation.\\
Affiliations:
{\em $^1$Laboratoire AIM, CEA/DSM-CNRS-Universit\'e Paris Diderot; IRFU/SAp, Centre de Saclay, 91191 Gif-sur-Yvette Cedex, France.
$^2$CNRS, IRAP, 14 avenue Edouard Belin, 31400 Toulouse, France.
$^3$Universit\'e de Toulouse, UPS-OMP, IRAP, 31400 Toulouse, France.
$^4$High Altitude Observatory, 3080 Center Green Drive, Boulder, CO, 80302 USA.
$^5$Universit\'e de Nice Sophia-Antipolis, CNRS, Observatoire de la C\^ote d'Azur, UMR 6202, BP 4229, 06304 Nice Cedex 4, France}}

\end{document}